\documentclass{article}
\usepackage{cite,spconf,amsmath,amssymb,amsfonts,graphicx,hyperref,multirow,makecell,threeparttable,booktabs}
\usepackage[table]{xcolor}
\usepackage{gensymb}

\title{
SOUNDCOMPASS: NAVIGATING TARGET SOUND EXTRACTION WITH \\
EFFECTIVE DIRECTIONAL CLUE INTEGRATION IN COMPLEX ACOUSTIC SCENES
}

\name{
Dayun Choi$^1$ and Jung-Woo Choi$^1$%
\sthanks{Corresponding author.
}
}
\address{
$^1$School of Electrical Engineering, KAIST, Daejeon, Republic of Korea \\
\{cdy3773, jwoo\}@kaist.ac.kr
}

\begin{document}
\ninept
\maketitle
\begin{abstract} 
Recent advances in target sound extraction (TSE) utilize directional clues derived from direction of arrival (DoA), which represent an inherent spatial property of sound available in any acoustic scene.
However, previous DoA-based methods rely on hand-crafted features or discrete encodings, which lose fine-grained spatial information and limit adaptability.
We propose SoundCompass, an effective directional clue integration framework centered on a Spectral Pairwise INteraction (SPIN) module that captures cross-channel spatial correlations in the complex spectrogram domain to preserve full spatial information in multichannel signals.
The input feature expressed in terms of spatial correlations is fused with a DoA clue represented as spherical harmonics (SH) encoding.
The fusion is carried out across overlapping frequency subbands, inheriting the benefits reported in the previous band-split architectures.
We also incorporate the iterative refinement strategy, chain-of-inference (CoI), in the TSE framework, which recursively fuses DoA with sound event activation estimated from the previous inference stage.
Experiments demonstrate that SoundCompass, combining SPIN, SH embedding, and CoI, robustly extracts target sources across diverse signal classes and spatial configurations.
\end{abstract}
\begin{keywords} 
directional clue, target sound extraction, spectral pairwise interaction, spherical harmonics, iterative refinement.
\end{keywords}
\section{Introduction} \label{sec_intro}
Target sound extraction (TSE) \cite{uss} refers to the task of selectively extracting a target audio source from a complex acoustic scene. TSE has gained increasing attention due to its wide range of practical applications in hearing aids \cite{hearing-aids}, augmented/virtual reality (AR/VR) \cite{vr-ar}, and teleconferencing \cite{teleconferencing}. In these scenarios, isolating a desired source from interfering signals and background noise is critical for both human perception and machine-based recognition.

Recent studies have investigated TSE using auxiliary clues that guide the model toward the target source, particularly in deep learning-based frameworks.
Illustrative auxiliary clues encompass class labels \cite{semantic-hearing, interaural}, text descriptions \cite{clapsep, language}, visual cues \cite{conceptbeam}, or their combinations \cite{multimodal}. In addition to these, the direction of arrival (DoA) has been utilized as a notable clue that leverages the spatial characteristic to isolate a target from interfering sources, irrespective of temporal or spectral attributes.

The effective use of DoA clues hinges not only on selecting input features that describe the spatial aspects of multichannel signals but also on the clue-fusion architecture in which DoA clues are articulated and integrated with these input features.
With respect to input features, prior work \cite{rezero, ssdq} has focused on manually designed features such as inter-channel phase differences (IPD) or inter-channel level differences (ILD), which improve TSE performance compared to using raw waveforms or complex spectrogram input. Nevertheless, they might lose essential spatial information, and whether such features are the optimal choice for capturing spatial relationships remains an unresolved issue.

Beyond the choice of input features, prior studies have also differed in how DoA clues are represented and fused with these features. Some studies \cite{rezero, ssdq, doa-or-speaker} used IPD and target phase difference (TPD) computed from target DoA and known microphone positions.
Other approaches \cite{dse, selective, soundtrc} adopted one-hot or binary encodings, which ignore the continuous and periodic nature of angular space by treating adjacent directions as independent categories, leading to increased input dimensionality and hindering generalization to unseen or intermediate directions.
In these approaches, DoA clues were combined with input features through operations such as multiplication \cite{dse}, initial recurrent states \cite{selective}, or attention keys/values \cite{soundtrc}.
More recently, M2M-TSE \cite{m2m-tse} and DSENet \cite{dsenet} employed cyclic positional (cyc-pos) embeddings that explicitly capture the periodic structure of angular space. M2M-TSE further applied them by broadcasting across time only when the target is active and multiplying them with input features. Since the exact duration of the target activity is generally unknown, this approach needs to handle temporal uncertainty when incorporating directional clues.

\begin{figure*}[t] \label{fig_model}
\centerline{\includegraphics[width=0.85\textwidth]{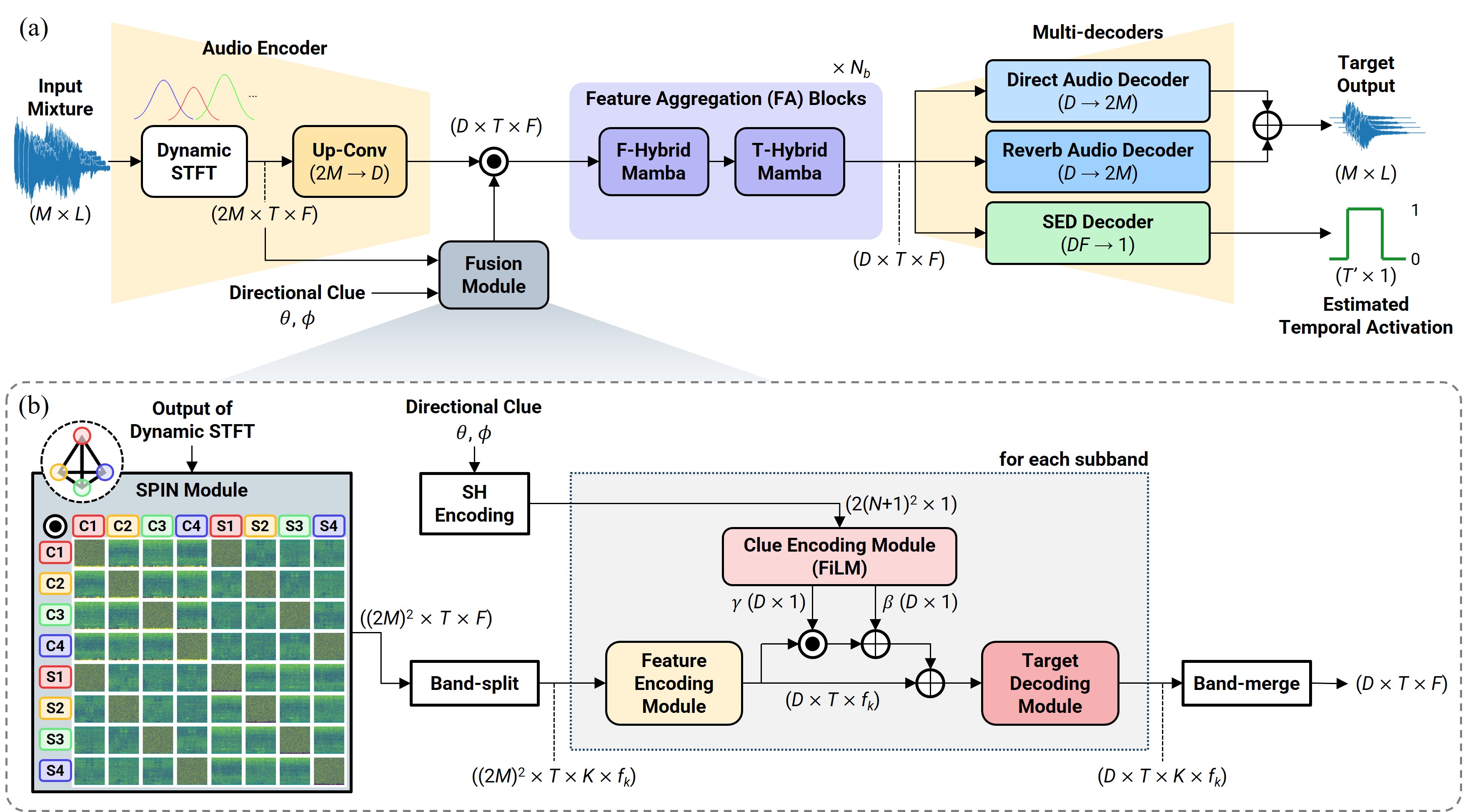}}
\vspace{-7pt}
\caption{(a) Overall architecture of SoundCompass for DoA-based target sound extraction and (b) details of a fusion module including a Spectral Pairwise INteraction (SPIN) module and integrating directional clue by feature-wise linear modulation (FiLM) for $K$ subbands.}
\vspace{-7pt}
\end{figure*}

To address these limitations, we propose SoundCompass, a framework for more effective directional clue integration.
The key contributions of the proposed framework are as follows:

\noindent \textbf{Spectral Pairwise INteraction (SPIN) input feature.} To build a general input feature capturing rich spatial information, we propose a Spectral Pairwise INteraction (SPIN) module that captures all pairwise interactions between sinusoidal components on complex spectrogram of multichannel signals. SPIN features are then fused with a DoA clue in overlapping frequency subbands to capture frequency-dependent spatial cues, extending the advantages of prior band-split approaches \cite{bsrnn, bandit, stereophonic, stem}. 

\noindent \textbf{Spherical harmonics (SH) embedding for DoA clues.} We employ spherical harmonics (SH) as the DoA clue embedding, which provides a continuous angular representation across the 2D sphere. The SH embedding enables the model to handle any DoA value without discretization.

\noindent \textbf{Iterative refinement with temporal clue.} Furthermore, we adopt an iterative refinement strategy inspired by the chain-of-inference (CoI) paradigm \cite{leveraging, self-guided, deepasa}, where the estimated temporal activation is recursively fused with the DoA clue to the subsequent stage, enabling the model to improve separation quality under challenging multi-source conditions progressively.
%
\section{SoundCompass Framework} \label{sec_method}

%
\subsection{Model Architecture} \label{subsec_model}
Our proposed model is based on DeepASA \cite{deepasa} backbone, which achieved state-of-the-art performance in universal source separation (USS) and sound event localization and detection (SELD) on their benchmarks. The architecture sequentially applies multi-head self-attention and Mamba feedforward networks along spectral and temporal dimensions separately to capture object-level features from mixtures. 
The original backbone separates object features without any clue, and our focus here is how to effectively guide the object separation process using a DoA clue.

The overall architecture for achieving this objective is illustrated in Fig.~\ref{fig_model}(a), excluding the batch dimension. A multichannel mixture is first transformed into a complex spectrogram of shape $2M \times T \times F$ using the short-time Fourier transform (STFT), where $M$, $T$, and $F$ denote the number of microphones, time frames, and frequency bins, respectively. Instead of a fixed window, a learnable Gaussian window parameterized by adaptive mean and standard deviation is applied, allowing each frame to be spectralized from different temporal focus and spread. The spectrogram is then mapped into a feature space, increasing the channel dimension from $2M$ to $D$ through a 2D convolutional encoder (kernel size 3 and stride 1) that extracts spatial cues and local spectral–temporal patterns. The resulting feature is modulated by a fusion module to align extracted features with the directional clue, as described in section \ref{subsec_clues}. Subsequently, fused features are processed by feature aggregation (FA) blocks that analyze spectral and temporal dependencies and separate features corresponding to the directional clue. Finally, two audio decoders reconstruct the multichannel direct sound and reverberation separately by reducing the channel dimension from $D$ back to $2M$ using 2D convolutional layers of kernel size 3 and stride 1, followed by an inverse STFT (iSTFT) to recover the target waveform.
\subsection{Directional Clue Integration} \label{subsec_clues}

\noindent \textbf{Spectral Pairwise INteraction (SPIN).} The fusion module takes the complex spectrogram and directional clues to generate the spatial feature mask that guides the DeepASA system to the target direction. To aid in extracting spatial details from the complex spectrogram, we present a Spectral Pairwise INteraction (SPIN) module. Within this module, 
the cosine and sine components derived from the phase of each multichannel complex STFT are multiplied pairwise across channels,
yielding a channel dimension of $(2M)^2$. This multiplication enhances the recognition of inter-channel phase or time differences, as well as level differences when necessary. The sinusoidal products, confined within a range of $\pm1$, ensure stable learning dynamics during training. Given that inter-channel relationships often vary by frequency, we adopt a band-split strategy \cite{bsrnn, bandit, stereophonic, stem} to ensure that inter-channel features are developed and merged with directional clues in each frequency band. Specifically, we employ overlapping subbands based on the 12-TET Western musical scale \cite{bandit, stem}, using $K=31$ subbands, with narrower bandwidths at lower frequencies and wider bandwidths at higher frequencies. This approach promotes continuity across overlapping frequencies and minimizes information loss at subband boundaries.

\noindent \textbf{Spherical harmonics (SH) embedding.} As an accurate and continuous representation of the DoA clue $(\theta, \phi)$, we employ spherical harmonics (SH) \cite{fundamentals} as embeddings, as depicted in Fig.~\ref{fig_model}(b). Unlike one-hot embeddings, SH embeddings allow for the representation of angles without the need for discretization. Additionally, in contrast to cyc-pos embeddings, which define azimuth and elevation angles separately, SH embeddings can describe the position on an $S^2$ sphere without coordinate separation, thus providing a consistent representation regardless of coordinate rotation. The complex spherical harmonics of order $n$ and degree $m$ are defined as
\begin{equation}
Y_n^m(\theta, \phi) =
\sqrt{\frac{(2n+1)}{4\pi} \frac{(n-m)!}{(n+m)!}}
P_n^m(\cos \theta)e^{i m \phi},
\end{equation}
where $P_n^m(\cdot)$ denotes associated Legendre functions. We use up to the 5-th order encoding by stacking the real and imaginary components of spherical harmonics, yielding an embedding vector of dimension $2(N+1)^2$ for $N=5$.

\noindent \textbf{Fusion in subbands.} The encoded SH vector is fused with the output from the SPIN module in each subband. The clue encoding module is a FiLM \cite{film} layer generating scale ($\gamma$) and shift ($\beta$) parameters for feature modulation, combined with a residual connection from the feature encoding module. This design enables fine-grained spatial conditioning without hand-crafted feature engineering. Each encoding and decoding block consists of a linear layer, adaptive layer normalization (AdaNorm) \cite{adanorm}, and a parametric rectified linear unit (PReLU) activation.

\begin{figure}[t]
\centerline{\includegraphics[width=\columnwidth]{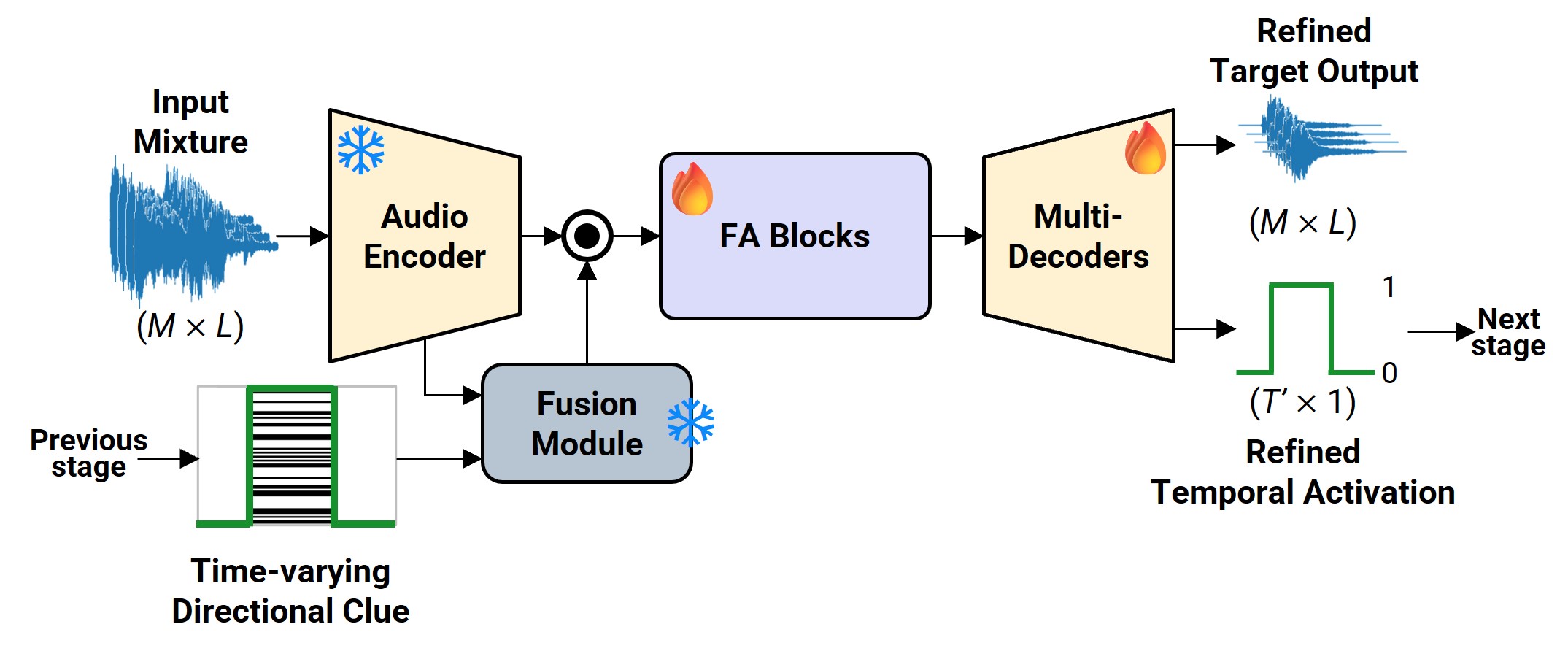}}
\vspace{-10pt}
\caption{Details of iterative refinement.}
\vspace{-8pt}
\label{fig_iter}
\end{figure}

\noindent \textbf{Iterative refinement.} To enhance robustness, we incorporate a sound event detection (SED) decoder following the on/off decoder structure of DeepASA \cite{deepasa}. This module outputs a frame-wise binary sequence that indicates the presence of the target source at each time step. As illustrated in Fig.~\ref{fig_iter}, this sequence is combined with the SH embedding to form a time-varying directional clue of shape $T' \times 2(N+1)^2$, where $T'$ denotes the sequence length of the SED decoder output. The clue is then linearly interpolated to $T$ in the time dimension and recursively injected into subsequent TSE stages. This chain-of-inference (CoI) strategy enables the model to iteratively refine its extraction by aligning directional information with temporal dynamics. 
During training, the already trained first-stage model is kept fixed, and only the subsequent stage is fine-tuned from the first-stage model using oracle time-varying clues while fixing its encoder and fusion module. This fine-tuned stage then functions as the subsequent stage. At evaluation, the complete pipeline is used: the output of the first-stage model is combined with the SH embedding and fed into the subsequent stage, which relies on SED predictions from the previous stage to progressively refine the separation of the target source.
\subsection{Loss Functions} \label{subsec_loss}
A linear combination of signal-to-noise ratio (SNR) and scale-invariant signal-to-noise ratio (SI-SNR) \cite{si-snr} loss with a ratio of 9:1 was used for the direct/reverb audio decoder and the sum of outputs from the decoders. Additionally, binary cross-entropy loss was employed for the SED decoder to estimate the temporal activation of the target source. All terms were summed with identical weights.
\section{Experiment and Analysis} \label{sec_experiment}
\subsection{Datasets} \label{subsec_dataset}
The proposed architecture was trained and evaluated on the Auditory Scene Analysis V2 (ASA2) dataset\footnote{\url{https://huggingface.co/datasets/donghoney22/ASA2_dataset}} \cite{deepasa}, which contains 13 audio classes with 2–5 foreground sources and one background noise per mixture. However, to align with our direction-based TSE model, we regenerated the dataset with stationary sources using the gpuRIR library\footnote{\url{https://github.com/DavidDiazGuerra/gpuRIR}} \cite{gpurir}, fixing each source at its initial position. Each mixture is 4 seconds long, sampled at 16 kHz, and recorded with a 4-channel tetrahedral microphone array of 4.2 cm radius centered in a cuboid room. The other configurations, including initial source positions and room reflections, follow those of the ASA2 dataset configuration. The resultant training, validation, and test sets comprise 50k, 2k, and 2k mixtures, respectively.

\subsection{Training Setups and Evaluation Metrics} \label{subsec_setup}
All training configurations largely followed \cite{deepasa}, with several modifications. Optimization was performed using AdamW with an initial learning rate of 0.0005, which was reduced by a factor of 0.1 if the validation loss did not decrease for five consecutive epochs. Gradient norm clipping was applied with a threshold of 5, and training was conducted for 100 epochs with a batch size of 2 on four GeForce RTX 4090 GPUs.

\begin{table*}[t]
\caption{Performance comparisons across models and structural variations of the proposed methods.}
\vspace{2pt}
\label{tab_result}
\centering
\resizebox{0.9\textwidth}{!}{
\begin{tabular}{lccccccc}

\toprule
Model & \multicolumn{2}{c}{SNR Metrics $\uparrow$} & \multicolumn{3}{c}{Spatial Errors $\downarrow$} & \multicolumn{2}{c}{Complexities $\downarrow$} \\
\cmidrule(lr){2-3} \cmidrule(lr){4-6} \cmidrule(lr){7-8} & SNRi (dB) & SI-SNRi (dB) & $\Delta$ILD (dB) & $\Delta$IPD (rad) & $\Delta$ITD ($\mu$s) & Param. & Mult-Adds \\

\midrule
\rowcolor{gray!15} \textit{Universal source separation} & & & & & & & \\
DeepASA \cite{deepasa} & 15.636 & 12.976 & 0.261 & 0.896 & 44.829 & 5.46 M & 74.85 G \\

\midrule
\rowcolor{gray!15} \textit{Target sound extraction} & & & & & & & \\
SSDQ (w. point spatial query) \cite{ssdq} & 5.949 & -1.171 & - & - & - & 3.91 M & 21.22 G \\
DSENet (w. cyc-pos ($\theta, \phi$)) \cite{dsenet} & 16.419 & 16.025 & - & - & - & 4.88 M & 86.89 G \\

\specialrule{0.3pt}{2pt}{2pt}
Proposed (DoA after FA) & 15.977 & 14.508 & 0.146 & 0.825 & 25.443 & 2.70 M & 20.49 G \\
\rowcolor{red!10} Proposed (DoA before FA) & 17.865 & 16.717 & 0.099 & 0.805 & 10.302 & 2.70 M & 20.49 G \\
\hspace{1em} remove an interaction in SPIN & 5.663 & 15.854 & 0.115 & 0.821 & 11.765 & 2.59 M & 20.49 G \\
\hspace{1em} replace SH to cyc-pos ($\theta, \phi$) & 17.696 & 16.538 & 0.100 & \textbf{0.782} & 12.747 & 2.70 M & 20.49 G \\
\hspace{1em} remove a band-split structure & 17.524 & 16.238 & 0.104 & 0.808 & 14.513 & 2.16 M & 20.49 G \\

\rowcolor{red!10} \hspace{1em} add an SED decoder & 17.884 & 16.780 & 0.098 & 0.800 & 9.993 & 4.09 M & 23.46 G \\
\rowcolor{red!20} \hspace{2em} refine iteratively ($\times 2$) & \textbf{18.196} & \textbf{17.079} & \textbf{0.093} & 0.789 & \textbf{9.714} & +3.48 M & +24.01 G \\
\bottomrule
\end{tabular}
}
\vspace{-3pt}
\end{table*}

The extraction performance was evaluated using SNR and SI-SNR improvements over the mixture. To further examine the spatial fidelity of multichannel extraction, the consistency of inter-channel cues was assessed by computing the mean absolute error (MAE) between estimation and ground truth, including $\Delta$ILD, $\Delta$IPD, and $\Delta$ITD across all microphone pairs. The ITD is derived using the generalized cross-correlation phase transform (GCC-PHAT)\footnote{\url{https://github.com/vb000/SemanticHearing}} \cite{semantic-hearing}. All the above metrics were computed for each source in each mixture and then averaged. In addition, the model complexity was evaluated in terms of the number of trainable parameters (Param.) and total multiplications and additions (Mult-Adds)\footnote{\url{https://github.com/TylerYep/torchinfo}}.
\subsection{Analysis of Results} \label{subsec_analysis}
Table \ref{tab_result} presents comparisons to other TSE systems, as well as ablation studies of the proposed SoundCompass framework. First, the vanilla DeepASA model \cite{deepasa}, representing universal source separation (USS) without any injected clue, achieves an SNRi of 15.6 dB and SI-SNRi of 13.0 dB. This unguided separation provides a baseline for evaluating the benefit of direction-aware extraction.
To evaluate the benefit of DoA clue injection, we compared early and late fusion strategies. Injecting DoA clues before feature aggregation (FA) blocks using the fusion module consistently improves performance over late integration (after FA). This confirms the importance of exploiting spatial cues at early stages to achieve better target sound extraction and spatial fidelity.

We then compared SoundCompass with recent DoA-based single-channel TSE baselines, SSDQ\footnote{\url{https://github.com/zhuxj821/SSDQ}} \cite{ssdq} and DSENet\footnote{\url{https://github.com/jingkangqi/DSENet}} \cite{dsenet}. For fair evaluation, we adapted their designs to match our setup of using one DoA clue: a point spatial query directing a point $(\theta, \phi)$ instead of a region query for SSDQ and extended cyc-pos embedding to elevation as well as azimuth without beamwidth control for DSENet, following their training configurations. SSDQ, which relies on hand-crafted features, performs poorly in multi-source mixtures, while DSENet shows noticeable improvements but still falls short of SoundCompass. Notably, our method achieves higher SNR metrics while maintaining lower computational complexity, indicating both effectiveness and efficiency.

\begin{figure}[t]
\centerline{\includegraphics[width=\columnwidth]{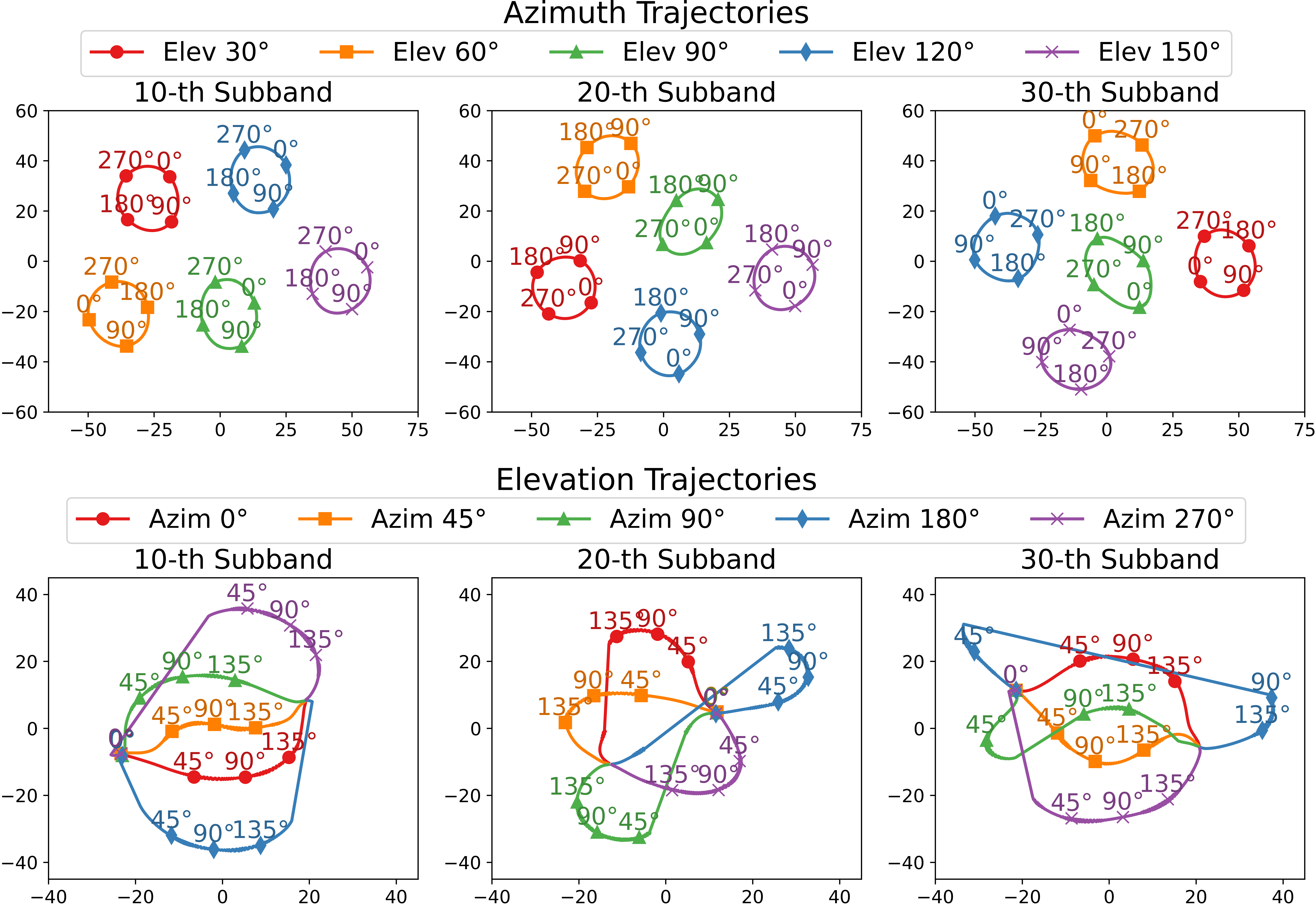}}
\vspace{-10pt}
\caption{The t-SNE trajectories of the FiLM scale ($\gamma$) parameters across three subbands,
with respect to azimuth (top, for 5 fixed elevations) and elevation (bottom, for 5 fixed azimuths).}
\vspace{-8pt}
\label{fig_tsne}
\end{figure}

\begin{figure}[t]
\centerline{\includegraphics[width=0.9\columnwidth]{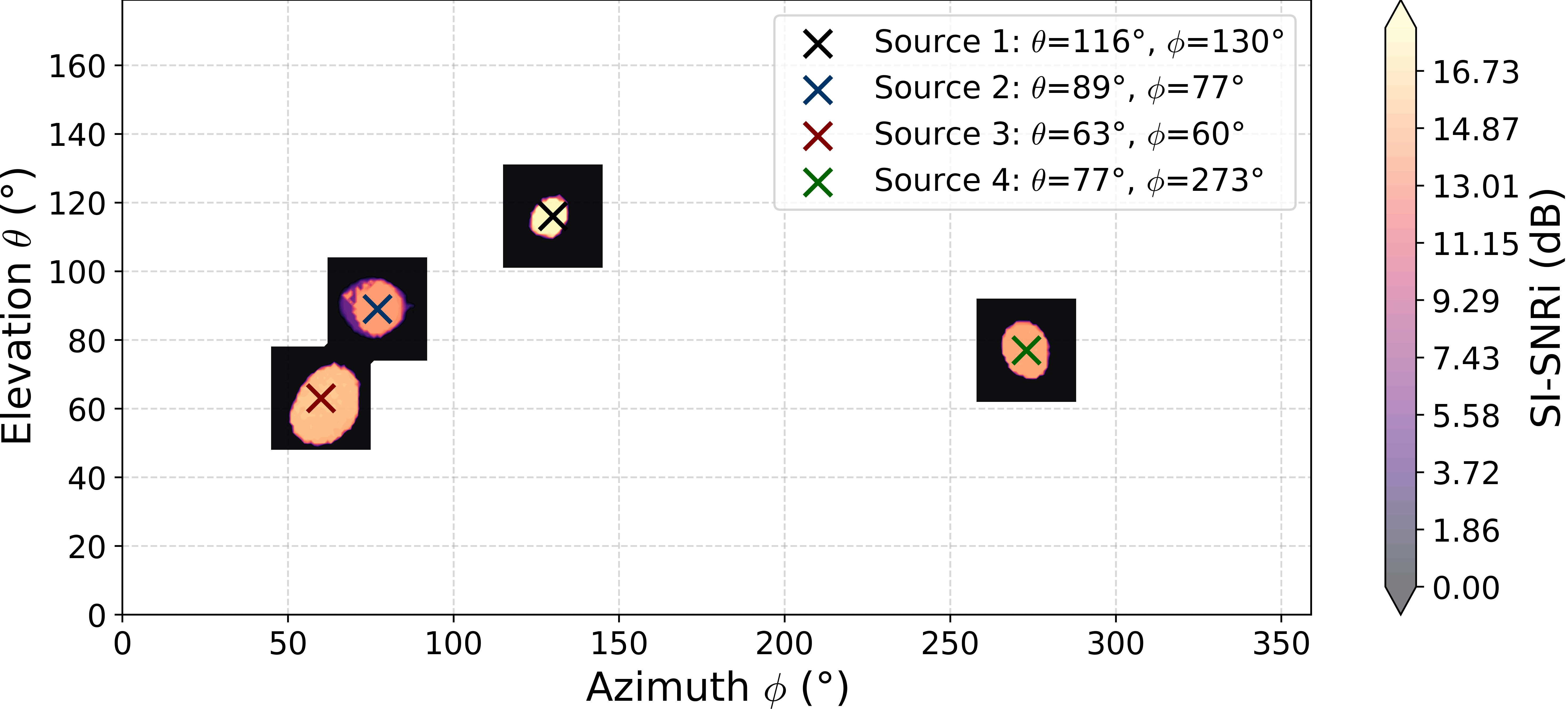}}
\vspace{-10pt}
\caption{An example of SI-SNRi contour maps within $\pm15^{\circ}$ from each target direction marked as ``X" in a cuboid room of size [width, length, height] = [5.57, 5.20, 3.79] m with an RT60 of 0.32 s.}
\vspace{-8pt}
\label{fig_contour}
\end{figure}

Ablation studies further highlight the contribution of each component. Removing pairwise interactions in the SPIN module (i.e., using only the raw $2M$ cosine and sine components without multiplication) causes a substantial degradation, highlighting the importance of cross-channel correlation modeling. Replacing spherical harmonics (SH) with cyc-pos embeddings slightly degrades performance, while eliminating the band-split structure also reduces accuracy, underscoring the importance of frequency-dependent spatial cues. Incorporating an SED decoder shows modest improvements, while iterative refinement further boosts performance, demonstrating the advantage of progressively refining activations at the cost of additional parameters. 

To better understand how the directional clue is embedded, Fig.~\ref{fig_tsne} visualizes the t-SNE trajectories of the FiLM scale ($\gamma$) parameters from the clue encoding module across three subbands. For 5 fixed elevations, we projected the scale parameters across azimuths into the same feature space using t-SNE; the same procedure was applied to visualize variations across elevations for 5 fixed azimuths. Azimuthal variations form near-circular manifolds that remain distinct across elevations, indicating that angular periodicity is preserved. In contrast, elevation trajectories evolve from $0^{\circ}$ to $180^{\circ}$ and converge to similar points across azimuths, reflecting the continuous nature of vertical cues. In addition, the different geometric patterns across subbands suggest that frequency-specific spatial correlations are captured, highlighting the benefit of band-split modulation.

Fig.~\ref{fig_contour} illustrates SI-SNRi sensitivity to deviations in directional clues. The contour maps show that performance peaks near the true source directions and degrades as the DoA estimate deviates by up to $\pm15^{\circ}$. Circular regions of high SI-SNRi form around each target position, demonstrating that SoundCompass framework effectively leverages directional guidance. Smaller regions indicate strong direction sensitivity, while broader regions suggest tolerance to some angular mismatch. This trade-off highlights the practical robustness of the proposed framework, as small DoA deviations are inevitable in real-world scenarios. The audio demo is available at \url{https://choishio.github.io/demo-SoundCompass/}.
\section{Conclusion}
We proposed SoundCompass, a DoA-based TSE framework that integrates spherical harmonics embedding with spectral pairwise interaction for efficient spatial conditioning. Through overlapping band-split modulation and sound event activation estimation, the model effectively captures both frequency-dependent and time-varying spatial cues with low complexity. Furthermore, the iterative refinement strategy highlights the advantage of coupling DoA clues with temporal dynamics, enabling a more robust extraction in diverse conditions. These results suggest promising directions toward more flexible source manipulation, such as handling dynamically moving sources by jointly estimating their time-varying DoA and activity.

\section{Acknowledgements}
This work was supported by the National Research Foundation of Korea (NRF) grant (No. RS-2024-00337945), the STEAM research grant (No. RS-2024-00464269) funded by the Ministry of Science and ICT of Korea government (MSIT), Hyundai Motor Company and Kia (No. G01250273), and the BK21 FOUR program through the NRF grant funded by the Ministry of Education of Korea government (MOE).


\bibliographystyle{IEEEbib}
\bibliography{references}

\end{document}